\documentclass[a4paper]{article}
\usepackage{graphicx,cite,a4wide,amssymb,amsmath}
\title{Correlated correlation functions in random-bond ferromagnets}
\author {
 \hfil Tom Davis${}^1$ and John Cardy${}^{1,2}$ \hfil \\ \\
 ${}^1$Theoretical Physics, Oxford University, \\
        1 Keble Road, Oxford OX1 3NP \\ \\
 ${}^2$All Souls' College, Oxford OX1 4AL
} 
\newcommand{\vac}{\left|\downarrow_1 \downarrow_2 \ldots \downarrow_n\right>}
\newcommand{\sg}{\sigma}
\newcommand{\ve}{\varepsilon}
\begin{document}
\maketitle
\begin{abstract}
The two-dimensional
random-bond $Q$-state Potts model is studied for $Q$ near 2 via the perturbative
renormalisation group to one loop. It is shown that weak disorder induces cross-correlations between
the quenched-averages of moments of the two-point spin/spin and energy/energy correlation functions,
which should be observable numerically in specific linear combinations of various quenched correlation
functions. The random-bond Ising model in $(2+\epsilon)$ dimensions is similarly treated. As a
byproduct, a simple method for deriving the scaling dimensions of all moments of the local energy
operator is presented.
\end{abstract}
\section{Introduction}
When studying disordered systems at or near their critical points, the notion of a single scaling
dimension governing the behaviour of an observable field at large distances becomes
inadequate. Randomness gives rise  to a broad probability distribution for the scaling dimension of
a two-point correlation function $\ln G / \ln r$, so the
quenched average of $n^{\rm th}$ moments of the two-point function will have a non-linear dependence
on $n$: 
\[ \overline{G^n(0,r)}\sim r^{-2x_n} {\rm\  with\ } x_n \ne nx_1 \] where the over-bar indicates the
average  over
quenched disorder. Such multiscaling behaviour has been both predicted analytically and observed
numerically in a wide range of systems, but we will primarily concern ourselves with the
random-bond $Q$-state Potts model in 2D (the random-bond Ising model in $(2+\epsilon)$ dimensions is
briefly considered in section \ref{ssec:ising}). For $Q$ near 2, the critical behaviour of  this system is
obtainable via an $\epsilon$-expansion about the pure Ising-model fixed point, with an expansion
parameter proportional to the specific heat exponent $\alpha$. (We recall the Harris criterion
\cite{har:cri}, which
states that bond randomness will be relevant if and only if the specific heat exponent is
positive: hence the disorder is marginal for the Ising model ($Q$=2) and relevant for all $Q$
greater than this value). Following initial work by Ludwig \cite{lud:inf,lud:cri}, multiscaling
behaviour of moments of the two-point spin-spin correlation function
has been well established in this model both theoretically and numerically (see e.g.,
\cite{lew:ss,ddp:rsb,jaccar:tm}), and multiscaling in the corrections to scaling for
the energy-energy correlator has been predicted. (The energy-sector calculation is complicated by
the fact that scaling operators correspond to irreducible representations of the permutation group
of replica indices, $S_n$).

In this paper we examine some further structure induced by the disorder, which shows up as
correlations between the two-point spin/spin and energy/energy 
functions. (In fact, the non-trivial replica structure of the energy sector means that a distinction
must be made between the connected and disconnected two-point energy correlators: these may have
different behaviour with the spin/spin function, and also be mutually correlated). For example, does
\[ \overline{\langle\sg(0)\sg(R)\rangle^p \langle\ve(0)\ve(R)\rangle^q}_{\!(c)}\sim
R^{-2(x_{\sg,p}+x_{\ve,q})} \] 
or is a more complex behaviour to be expected? We show that this is dependent on both the model in
question, and whether the connected or full $\left<\ve\ve\right>$ correlator is taken. 

In section \ref{sec:intro} we shall introduce the $Q$-state Potts model and the replica method used
to carry out the average over quenched disorder, before describing results already obtained for this
model. In section \ref{ssec:energy} we present calculations of scaling dimensions (to one loop)
of moments of the energy-operator, and then show (section \ref{ssec:mixed})that a non-trivial
behaviour of the mixed spin-energy moments is 
to be expected. This indicates that the quantities $\ln G_{\sg\sg}/\ln R$ and
$\ln G_{\ve\ve}/\ln R$ should be considered as being drawn from a joint probability
distribution function that is not equal to the product of the individual marginal distributions,
i.e., the two quantities are not statistically independent. We then make contact with
quantities available via numerical calculations, by expressing two-point functions of the irrep
scaling operators as linear combinations of  various quenched averaged energy correlation functions
(section \ref{ssec:rstruc}). Finally we describe a similar calculation for the random-bond Ising
model in $(2+\epsilon)$ dimensions (section \ref{ssec:ising}).

\section{Model and previous results}\label{sec:intro}
We write the reduced lattice hamiltonian for the $Q$-state Potts model with weak bond disorder as
 
\[ H=\sum_{\left<i,j\right>}\frac{(-J_0+\delta J_{i,j})}{kT}\delta_{s_i,s_j} \]
where $s_i\in\left\{1\ldots Q\right\}$ are the Potts spin variables, $J_0$ is the average bond
strength and $\delta J_{i,j} \ll J_0$ are the local fluctuations about $J_0$, assumed to be
completely uncorrelated. The sum is taken over all nearest-neighbour pairs. Taking the local energy
density \[ e(r)=e_{\left< i,j \right>}=((-J_0+\delta J)/kT_c)\delta_{s_i,s_j} \] and using replicas
to average over the 
disorder, we end up with an effective hamiltonian in terms of $n$ replicas of the system coupled
together. We assume that $n$ is large, and take the $n\to 0$ limit to perform the quenched
average at the end of the calculation (see \cite{lub:rep} or \cite[chapter 8]{car:sca} for an
overview of the replica method in field-theoretic calculations): 
\[ H_{\rm eff}=\sum_{a=1}^n \left( H^*_a (Q)+t\sum_{r} e_a(r)\right)-\sum_{r} \Delta\sum_{a,b}^n
e_a(r)e_b(r) \]
 where $H^*(Q)$ is the reduced hamiltonian for the critical pure Potts model, $t$ is the reduced
temperature $(T-T_c)/T_c$ and $\Delta$ is proportional to the second cumulant of the distribution for
disorder (power-counting arguments show that higher cumulants are irrelevant at the pure fixed point
for $Q$ near 2). We note further that the expectation value of $e$ only shifts the
critical temperature and cumulants, without affecting the critical exponents in any way, so we can
write $e(r)=\left<e\right>+\ve(r)$ and absorb the first term into the definitions of the other
parameters. Also, terms with $a=b$ in the double sum can also be neglected since they are either
irrelevant 
by power-counting or yield disconnected diagrams which do not contribute to the renormalisation
\cite[appendix B]{lud:cri}. Finally, moving to the continuum limit, we find

\[ H_{\rm eff}=\sum_{a=1}^n H^*_a(q)+t\int d^2r\sum_{a=1}^n \ve_a(r)-\Delta\int d^2r \sum_{a\ne
b}\ve_a(r)\ve_b(r). \]
Correlation functions calculated against this effective, `replicated' hamiltonian will correspond to
correlators averaged against the initial hamiltonian with quenched disorder:
\begin{eqnarray*}
\overline{\langle\ve(r)\rangle_H}  & \longleftrightarrow &
\lim_{n\to0}\langle\ve_i(r)\rangle_{\rm rep} \\ 
\overline{\langle\ve(0)\ve(r)\rangle_H} & \longleftrightarrow &
\lim_{n\to0}\langle\ve_i(0)\ve_i(r)\rangle_{\rm rep} \\
\overline{\langle\ve(0)\rangle_H\langle\ve(r)\rangle_H} & \longleftrightarrow &
\lim_{n\to0}\langle\ve_i(0)\ve_j(r)\rangle_{\rm rep},\ i\ne j
\end{eqnarray*}
where $i$ and $j$ label the replicas the operators are lying in. Note that it is assumed that the
above correlators are independent of which replicas are actually taken --- this is the assumption of
replica symmetry, which appears to be valid for weakly disordered ferromagnets.

The quenched average of $p^{\rm th}$ moments of the spin operator was calculated by Ludwig for all
$p$, to one loop. The 
scaling dimension of the operator $\sg_1 \sg_2 \ldots \sg_p$ was found to be
\[ X_p=p x_\sg - \frac{y}{16}p(p-1) + O(y^2)\]
where $x_\sg$ is the scaling dimension of the spin operator at the pure fixed point and
 $y$ is the RG eigenvalue of $\Delta$, vanishing proportional to $(Q-2)$. For typical realisations of
disorder, the  behaviour of the two-point function at large $r$ is governed by a multifractal
exponent (c.f. \cite{hal:mf}). This is given by the saddle point of the Legendre
transformation of $X_p$, and was found to be  
 \[\alpha_0\equiv\left.\frac{\partial X_p}{\partial p}\right|_{p=0}\!\!=x_\sg + y/16 +O(y^2).\]

In general a given moment of the energy operators does not have a pure scaling behaviour. In the
replica formalism, our perturbing operator $\Delta \sum_{a\ne b}\ve_a\ve_b$ is a singlet under the
group of permutations of the replica indices, $S_n$ (in this context the permutation
group is often named the `replica' group instead). Thus scaling dimensions of operators will only
necessarily be constant on subspaces of operators corresponding to irreducible representations
(irreps) of
$S_n$. Degeneracies between irreps may arise --- in fact all irreps of moments of the spin operator are
degenerate, due to their different fusion rules with the perturbing operator  ---
but in general a given operator $\ve_1\ve_2\ldots\ve_q$ will have a `sum-of-powers' scaling
behaviour
\[ \left< \ve_1\ve_2\ldots\ve_q(0)\ \ve_1\ve_2\ldots\ve_q(r)\right> \sim \sum_{\mu\in{\cal
R}}A_\mu r^{-2x_\mu}\] 
 where $\mu$ runs over the different irreps of $S_n$ present in the decomposition of
$\ve_1\ve_2\ldots\ve_q$. 

Calculations for $q\le 3$ were given in \cite{lud:inf}. In each case the most 
antisymmetric irrep was the most relevant, and the scaling dimensions of these most relevant irreps were
linear in $q$, suggesting no multiscaling behaviour to this order. A more complex structure was
however present in the corrections to scaling.

\section{Calculation}\label{sec:calc}
\subsection{Energy Sector: irreducible representations of $S_n$ and $SU(2)$}\label{ssec:energy}
First we shall exploit the connection between the irreps.\ of $S_N$ and $SU(2)$ to obtain complete
sets of scaling dimensions for moments of the energy-energy 
correlation function, $\overline{\langle\ve(0)\ve(r)\rangle^q}$,  to one loop. At this order,
the effect of a shift in the short-distance cutoff on the couplings is most easily treated in terms
of the operator product expansion (OPE) (see e.g., \cite[chapter 5]{car:sca}). Consider a general
partition function $Z$ for a critical 
hamiltonian $H^*$, perturbed by a set of scaling fields $\phi_i$ with corresponding couplings $g_i$,
\[Z  = {\rm Tr}\ e^{-H^*-\sum_i g_i \int d^d r\,a^{x_i}\phi_i (r)}, \]
where $a$ is the microscopic cutoff. The couplings flow (to first order) as
\[ dg_k/dl=(d-x_k)g_k-\sum_{ij}c_{ijk}g_i g_j + \ldots \]
where $c_{ijk}$ are the OPE coefficients
\[\phi_i(r_1) \cdot \phi_j(r_2)\sim\sum_k \frac{c_{ijk}}{\left|r_1-r_2\right|^{x_i+x_j-x_k}}
\phi_k\left(\frac{r_1+r_2}{2}\right). \] 
 
Before the randomness is introduced, the operators $\ve_c \ve_d \ldots$
with $q$ distinct replica labels are degenerate. 
We need
to calculate the OPE coefficients of the disorder operator ${\cal O}=\sum_{a\ne b}\ve_a\ve_b$ within
this subspace.
In $SU(2)$ language let us denote the 
presence of an energy operator in replica $i$ by $\left|\uparrow_i\right>$ and its absence by
$\left|\downarrow_i\right>$. We can write a general operator by the action of a series of $SU(2)$
raising operators on a `vacuum' $\vac$:
\begin{gather*}
\ve_a \ve_b\ldots \longleftrightarrow \tau_a^+ \tau_b^+ \ldots \vac
\intertext{with}
q=\frac{1}{2}\sum_{i=1}^{n}(1+2\tau_i^z)=\frac{1}{2}n+\sum_i \tau_i^z.
\end{gather*}
In the subspace corresponding to the energy sector, the OPE of one of these operators with the
disorder operator ${\cal O}$ is equivalent to the action of the matrix
\[ {\cal M}=2\sum_{i \ne j}^{n}\tau_i^+\tau_j^{-} \]
on the corresponding state, with
$\left[ {\cal M}, \sum_i \tau_i^{z} \right]=0$ as we would expect. (The extra
factor of two arises from the two different ways of performing the operator contraction). If we define
the total spin vector $\vec S=\sum_{i=1}^n \vec \tau_i$ then we can write
\begin{eqnarray*}
{\cal M} & = & 2\sum_{i \ne j}^{n}\tau_i^{+}\tau_j^{-} \\
         & = & 2 \left( \sum_i \tau_i^{+}\right)\left(\sum_j \tau_j^{-}\right) -2\sum_i \tau_i^+
                    \tau_i^-  \\
         & = & 2( S^+ S^- -q) \\
         & = & 2(\vec S ^2-(S^z)^2+S^z-q)
\end{eqnarray*}
Using $q=S^z+\frac{n}{2}$ we can thus write
\[ {\cal M}=2\left(S(S+1)-(S^z)^2 - \frac{n}{2}\right). \]
However, we now have the problem of interpreting the
total spin angular momentum $S$. One solution is to ask which values of $S$ are consistent with the
known value of $S^z$ --- clearly, the possible values are
\[ S=\frac{n}{2},\, \frac{n}{2}-1,\,\frac{n}{2}-2,\,\ldots,\,\frac{n}{2}-q. \] We can support this by
comparing the irreducible representations of $SU(2)$ with the irreducible representations of $S_n$
considered by Ludwig. In the $S_n$ case, we have to consider the representations given by all Young
tableaux of $n$ boxes and at most two rows, with up to $q$ boxes in the second row. E.g., for $q=2$,
$n=6$ the possible irreps are \cite{ham:grp}

\begin{picture}(93,35)
\multiput(0,15.5)(15.5,0){2}{\framebox(15,15){$\ve$}}
\multiput(31,15.5)(15.5,0){4}{\framebox(15,15){\ }}
\end{picture}
\hfil
\begin{picture}(77.5,35)
\multiput(0,15.5)(15.5,0){2}{\framebox(15,15){$\ve$}}
\multiput(31,15.5)(15.5,0){3}{\framebox(15,15){\ }}
\put(0,0){\framebox(15,15){\ }}
\end{picture}
\hfil
\begin{picture}(62,35)
\multiput(0,15.5)(15.5,0){2}{\framebox(15,15){$\ve$}}
\multiput(31,15.5)(15.5,0){2}{\framebox(15,15){\ }}
\multiput(0,0)(15.5,0){2}{\framebox(15,15){\ }}
\end{picture}

These irreps
can be placed in one-to-one correspondence with the representations formed by the tensor product of
$n$ spin-half representations of $SU(2)$ with precisely $q$ of them spin-up. Of these latter
tableaux, a tableau with $(n-r)$ boxes in the first row and $r$ boxes in the second row has a
dimensionality of $D=n-2r+1$ corresponding to a total spin of $S=\frac{D-1}{2}=\frac{n}{2}-r$, as
was proposed by the consistency condition. Hence, in general the $q^{\rm th}$ moment of the energy
operator will have a series of irreps ${\mathfrak I}_{q,r}$ with scaling dimensions
\[ X_{q,r}(n)=q x_\ve -
y\cdot\frac{1}{4}\cdot2\left(\left(\frac{n}{2}-r\right)\left(\frac{n}{2}-r+1\right)-
\left(\frac{n}{2}-q\right)^2 - \frac{n}{2}\right)+O(y^2) \]
where  $r\in\{0,1,\ldots,q\}$, $y$ is the RG eigenvalue of the coupling to disorder and $x_\ve$ is the
scaling dimension of the energy operator at the pure fixed point,
\[ x_\ve=1-\frac{y}{2}+ O(y^2). \]
The factor of $\frac{1}{4}$ arises from the position of the disordered fixed point, set (to one
loop) by
\[ d\Delta/dl=\alpha\Delta+4(n-2)\Delta^2+O(\Delta^3). \]

After taking the $n \rightarrow 0$ limit we obtain
\[ X_{q,r}(0)=q - \frac{y}{2}\left(r(r-1)-q^2+q\right) +  O(y^2). \]
The most relevant scaling dimension  belongs to the  most antisymmetric irrep $\mathfrak{I}_{q,q}$
and has a linear dependence on  $q$. Note that the irreps $\mathfrak{I}_{q,0}$ and
$\mathfrak{I}_{q,1}$ are always degenerate at $n=0$. This must happen
because for $n\not=0$ they
have degeneracies 1 and $(n-1)$ respectively, while as $n\to0$ all
operators must arrange themselves to have  degeneracies $\propto n$, so
that the torus partition function is $1+O(n)$. In fact, this collision
of the scaling dimensions at $n=0$ will give rise to a logarithmic
operator \cite{JClog}.

The operators $\ve_1\ldots\ve_q$ span a vector space of dimension
$\binom{n}{q}$.  In general, a Young tableaux of shape $[n-r, r]$
corresponds  to a vector space of
dimensionality $\frac{(n-r)!(n-2r+1)}{(n-r+1)}$, and it can be shown that the sets of permitted
irreps ($r\in [0,q]$) exhaust this vector space. 
The dimensionality of a given irrep does not depend on the number of energy operators in the
tableau, and is preserved under the addition of  extra $\ve$-operators  via $S^+$
--- irreps related via the action of $S^\pm$ are joined in the plot below (figure
\ref{fig:energyspec}): 

\begin{figure}[htp]
\centering
\includegraphics[width=10cm, height=8cm]{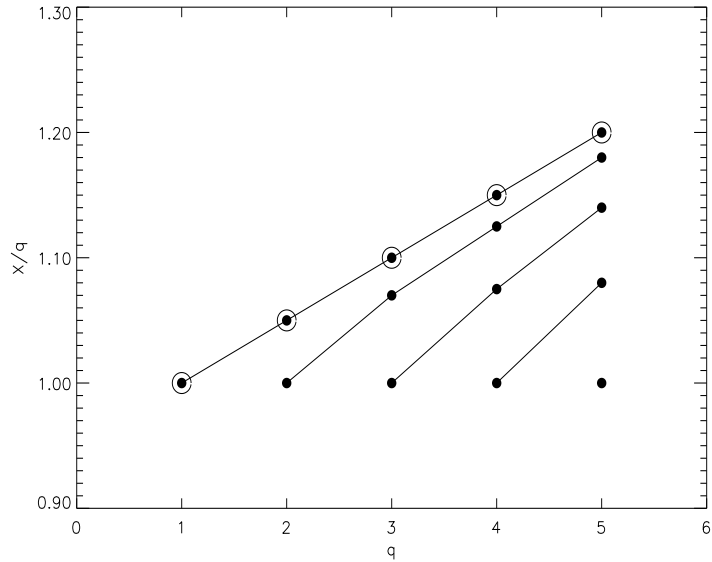}
\caption{Scaling dimensions of $q^{\rm th}$ moments of the energy operator to one loop at
$y=0.1$}\label{fig:energyspec} 
\end{figure}

The two-point function of the most antisymmetric irrep for a given $q$, $\mathfrak{I}_{q,q}$,
corresponds to the  
quenched average of the $q^{\rm th}$ moment of the connected correlation function,
\[ \overline{G_c^q}=\overline{(\ve_1-\ve_2)\ldots(\ve_{2q-1}-\ve_{2q})(0)
(\ve_1-\ve_2)\ldots(\ve_{2q-1}-\ve_{2q})(r)}\,. \]
This can be seen by noting that the state corresponding to the operator
$(\ve_1-\ve_2)\ldots(\ve_{2q-1}-\ve_{2q})$ is annihilated by the step-down operator
$S^-=\sum_{i=1}^{n}\tau_i^-$:
\[
\left(\sum_{i=1}^{n}\tau_i^-\right)(\tau_1^+-\tau_2^+)\ldots(\tau_{2q-1}^+-\tau_{2q}^+)\vac=0.
\]
All $\tau_i^-$ with $i>2q$ give zero when acting on everything to their right, so we only need
consider those $\tau_i^-$ with $i\le2q$. Take the terms of the step-down operator in pairs: the
first pair $(\tau_1^- + \tau_2^-)$ will act on everything to the right to give
\[(\tau_1^-\tau_1^+-\tau_2^-\tau_2^+)(\tau_3^+-\tau_4^+)\ldots(\tau_{2q-1}^+-\tau_{2q}^+)\vac=0.\]
The action of all other pairs similarly vanishes, so the state corresponding to the connected correlation
function is annihilated by $S^-$ and is therefore also the state corresponding to the most
antisymmetric irrep. 

\subsection{Mixed Sector: Exclusion argument}\label{ssec:mixed}
We now wish to consider the behaviour of cross-moments, i.e., investigate the renormalisation of the
operators $\sg_1\ldots\sg_p\ve_1\ldots\ve_q$. Fortunately, given the behaviour of the
$\ve_1\ldots\ve_q$ operators 
considered above this is not a particularly difficult extension. To one loop, we again consider the
effect of the disorder operator $\sum_{a\ne b}\ve_a \ve_b$ on the operator $\sg_{i_1} \ldots \sg_{i_p}
\,\ve_{j_1}\ldots \ve_{j_q}$, and look for operators of the same form to be produced by all possible
contractions. These contractions are given by the fusion rules of our theory at the pure fixed
point:
\begin{eqnarray*}
\ve_a \cdot \ve_b & \sim &\delta_{a,b} \\ 
\sg_a \cdot \sg_b &\sim &\delta_{a,b}\left(1+\left(\frac{1}{2}+O(Q-2)\right)\ve_a\right)
\end{eqnarray*}

Contractions are only possible within a given replica.
Given the fusion rules, at this order there are only two possibilities:
\begin{enumerate}
\item $a \in \{j\}, b \notin \{i\}\cup \{j\}$: one of the energy operators in the disorder operator
contracts against another energy 
operator $(\ve\ve\to 1)$, the other lies in an `empty' replica. This  gives the same behaviour as in
the pure energy 
sector, but with a shift in the effective 
number of replica indices: $n\rightarrow n-p$
\item $a,b\in \{i\},\, a\ne b$: both energy operators in the disorder operator contract with spin
operators, 
giving rise to further spin operators. As considered by Ludwig, this gives rise to a term
$\frac{1}{4}p(p-1)$
\end{enumerate}
Any other attempted contractions will evaluate to zero when the the trace over the fields is taken.
Substituting these changes into our expression for the energy sector, after taking the $n\to0$ limit
we find 
\[ X_{p,q,r}=p x_\sg + q - \frac{y}{4}
 \left[2\left\{ \left( \frac{p}{2}+r\right)\left(\frac{p}{2}+r-1\right)-\left(q+\frac{p}{2}\right)^2
+ \frac{p}{2}+q\right\} + \frac{1}{4}\left(p^2-p\right)\right]+O(y^2) \]
 where as before $r\in[\,0,q\,]$ and $x_\sg$ is the scaling dimensions of the spin
operator at the pure fixed point. The spin operators exclude the
energy operators from their replicas, giving rise to an effective shift in the number of replicas
$n\to n-p$.

 The scaling dimensions of the most antisymmetric irreps $r=q$ are unaffected by this
exclusion process, as they have no $n$-dependence. This means that,
to one loop, there will be no extra structure
in the quenched average $\overline{\left<\sg(0)\sg(R)\right>\left<\ve(0)\ve(R)\right>_c}$. In
section \ref{ssec:ising}, however, we shall demonstrate that this quantity does have extra structure
for the case of the random-bond Ising model in $2+\epsilon$ dimensions. 

By analogy with the $SU(2)$ argument, it is possible to consider the mixed sector by writing our
operators $\{\ve, \sg, 1\}$ as a fundamental representation of $SU(3)$. This analysis leads to the same
results, as the $\{1,\ve\}$ $SU(2)$ subgroup effectively decouples from the spin operator.

As one check on these results, standard arguments in probability theory (see, e.g.,\cite{fel:prob})
require that 
$-X_{p,q,q}$ should be convex for all increases in $p$ or $q$, both individually and jointly.
Considering the second derivatives of this quantity w.r.t.{} $p$ and $q$, this requirement can be seen
to be trivially satisfied for $y>0$. 

\subsection{Replica Structure}\label{ssec:rstruc}
Given that non-trivial structure in the spin-energy sector only appears to sub-leading order, it is useful
to be able to write two-point functions of irreducible representations of the replica group in terms
of numerically available quantities. We have already noted that the most relevant, most
antisymmetric irreps correspond to moments of the connected part of the two-point correlation
function (section \ref{ssec:mixed}) and would like to be able to say something similar for a general
irrep. In Appendix A we detail the combinatorics needed to decompose a two-point function of irreps
into a linear combination of various other two-point functions: here we shall merely note a few
results.

We find explicitly that the most antisymmetric irrep at a given $q$, $\mathfrak{I}_{q,q}$ has a
two-point function
\begin{eqnarray*}
 \langle\mathfrak{I}_{q,q}(0)\mathfrak{I}_{q,q}(R)\rangle & \sim &
\langle(\ve_1-\ve_2)\ldots(\ve_{2q-1}-\ve_{2q})(0)\,
(\ve_1-\ve_2)\ldots(\ve_{2q-1}-\ve_{2q})(R) \rangle_{\rm rep} \\
 &\sim& \overline{G_c^q(R)}
\end{eqnarray*}

In the $q=1$ case, both the possible irreps $\mathfrak{I}_{1,0}$ and $\mathfrak{I}_{1,1}$ correspond
to the connected correlator $\overline{\langle\ve(0)\ve(R)\rangle}-
\overline{\langle\ve(0)\rangle\langle\ve(R)\rangle}$. This is in fact a general feature, since the lowest
two irreps for arbitrary $q$ are always degenerate after taking the $n\to0$
limit.

As an example of a correlation function which has a non-trivial behaviour on addition of $p$
spin-spin correlators, the irrep $\mathfrak{I}_{2,1}$  is a non-leading irrep
which can be written as
\[\langle\mathfrak{I}_{2,1}(0)\mathfrak{I}_{2,1}(R)\rangle
\sim\overline{\langle \ve(0)\ve(R)\rangle\langle\ve(0)\ve(R)\rangle} -
4\overline{\langle\ve(0)\ve(R)\rangle \langle\ve(0)\rangle\langle\ve(R)\rangle}
+3\overline{\langle\ve(0)\rangle\langle\ve(0)\rangle\langle\ve(R)\rangle\langle\ve(R)\rangle}. \]
If we take a factor of $\langle\sg(0)\sg(R)\rangle^p$ inside each quenched average on the
RHS, the resulting scaling dimension will be
\begin{gather*}
 X_{p,2,1}  =  px_\sg+2x_\ve-\frac{y}{4}(-8-2p+\frac{1}{4}(p^2-p))+O(y^2) \\
 \ne \\
X_{p,0,0}+X_{0,2,1} =  px_\sg+2x_\ve-\frac{y}{4}(-8+\frac{1}{4}(p^2-p))+O(y^2)
\end{gather*}
Thus this correlation function directly exhibits the cross-structure between the two-point function of
the spin operator and the full two-point function of the energy operator.

\subsection{Ising model in $2+\epsilon$ dimensions}\label{ssec:ising}
As an alternative to perturbing the pure Ising model by changing the number of possible values for
the spin, we can consider the Ising model in $2+\epsilon$ dimensions. The specific heat exponent
$\alpha$ is zero at $d=2$ and small for $d=3$ ($<0.1$), so we shall assume that
$\alpha=O(\epsilon)$ with $\epsilon \ll 1$. The bond disorder is relevant for all $\epsilon>0$, and
we assume that the system will flow to a nearby disordered fixed point. The calculation of scaling
dimensions at this fixed point is very similar to that for the disordered Potts model --- the only
difference is that the OPE rules are altered to
\begin{eqnarray*}
 \ve_a \cdot \ve_b & \sim & \delta_{a,b}(1 + c \ve) \\
 \sg_a \cdot \sg_b & \sim & \delta_{a,b}(1 + \big((1/2)+O(\epsilon)\big)\ve)
\end{eqnarray*}
The OPE coefficient $c$ is zero for $d=2$ due to the self-duality of the Ising model in two
dimensions \cite[chapter 8]{car:sca}, but non-zero for $d>2$. Repeating the analysis of section
\ref{ssec:mixed} we can obtain two more types of contraction between the disorder operator
$\sum_{a\ne b}\ve_a\ve_b$ and $\sg^p\ve^q$:
\begin{enumerate}
\item Both energy operators in the disorder operator contract with energy operators, giving rise to
two further energy operators in the same replicas. This gives a term $c^2 q(q-1)$.
\item The disorder operator contracts with one energy operator and one spin operator, giving rise to
one operator of each type with unchanged replica indices. This produces a term
$2\cdot(1/2)\cdot c\cdot p q=cpq$.
\end{enumerate}
The non-zero $c$ will shift the position of the fixed points, via the RG equation for the disorder
\[d\Delta/dl=\alpha\Delta+\left(4(n-2)+2c^2\right)\Delta^2+O(\Delta^3, \epsilon\Delta^2) \]
so we obtain scaling dimensions
\begin{multline*}
 X_{p,q,r}=p x_\sg + qx_\ve + \frac{y}{c^2-4}
 \biggl\{2\biggl[ \left( \frac{p}{2}+r\right)\left(\frac{p}{2}+r-1\right)-\left(q+\frac{p}{2}\right)^2
+ \frac{p}{2}\biggr] +\cdots \\
  \cdots+\frac{1}{4}\left(p^2-p\right)+c^2(q^2-q)+cpq\biggr\}+O(y^2) 
\end{multline*}

The $cpq$ term indicates that, with a non-zero $c$, we now have cross-structure between the 
two-point function of the spin operator and the {\it connected} two-point function of the energy
operator, i.e. 
$\overline{\left<\sg(0)\sg(R)\right>^p\left<\ve(0)\ve(R)\right>^q_c}\not=
\overline{\left<\sg(0)\sg(R)\right>^p}\cdot\overline{\left<\ve(0)\ve(R)\right>^q_c}$

\section{Conclusions}\label{sec:conc}
To one loop, the interaction between the spin and energy operators is an `exclusion effect':
the spin operators block replicas, shifting the $n$-dependence of the scaling dimensions of the
different irreps in the energy sector. For the Potts model this does not have any effect on the leading
behaviour, as the most relevant irrep irrep $r=q$ has a scaling dimension with no
$n$-dependence. The sub-leading terms are affected, however, and 
this may be picked up in numerical studies. By writing the irreps in terms of correlation functions
we have shown how linear combinations of correlators could be used to demonstrate the existence of
an underlying joint distribution function for the quenched average of spin/spin and energy/energy
moments. In the Ising model in $(2+\epsilon)$ dimensions the leading behaviour of mixed moments is
affected non-trivially. Also, three- or higher- point functions should also exhibit this
cross-sector behaviour, presumably with selection rules coming from the group structure of the
energy sector.

There remains the question of what happens when these calculations are continued to two loops. {\it
A priori}, there seems no reason to expect that the scaling dimensions of the most antisymmetric
irreps will continue to be
protected against corrections from cross-correlations between the spin and energy two-point
functions. We will content ourselves here with noting that a coulomb-gas
calculation along the lines of \cite{ddp:rsb} would give two-loop corrections to the scaling
dimensions of the operator $\sg\ve$ proportional to
\[ \int d^2y\,\langle \ve(0)\ve(y)\ve(1)\ve(\infty)\rangle\,\langle\ve(0)\ve(y)\sg(1)\sg(\infty)\rangle
\]
which we would expect to be non-zero, although a detailed calculation would of course be necessary
to confirm this.

\section{Acknowledgements}
 After this work was substantially completed, we heard of work by Jeng and Ludwig\cite{jenlud:rdl}
who, in the course of their investigation of random defect lines in 2D systems, also derived (by an
alternative method) the scaling dimensions of moments of the energy operator,
to two-loop order. We would like to thank them for interesting discussions on this and other
points. We also thank  A.~Cavagna and R.~Stinchcombe for useful conversations. 
This work was supported in part by the Engineering and Physical Sciences
Research Council under Grant GR/J78327 and Studentship 98311143.
\newpage
\appendix
\section{Decomposition of $\langle\mathfrak{I}_{q,r}(0)\mathfrak{I}_{q,r}(R)\rangle$ into quenched
averaged energy correlators} 
Start with the most antisymmetric irrep for $q$ energy operators, $\mathfrak{I}_{q,q}$, corresponding
to the Young tableau

\hfil
\begin{picture}(123,35)
\put(0,15.5){\framebox(15,15){$\ve_1$}}
\put(15.5,15.5){\framebox(30,15){$\ldots$}}
\put(46,15.5){\framebox(15,15){$\ve_q$}}
\put(61.5,15.5){$\underbrace{\framebox(60,15){$\ldots$}}_{n-2q}$}
\put(0,0){\framebox(15,15){}}
\put(15.5,0){\framebox(30,15){$\ldots$}}
\put(46,0){\framebox(15,15){}}
\end{picture}
\hfil

The correlator $\langle\mathfrak{I}_{q,q}(0)\mathfrak{I}_{q,q}(R)\rangle$ is  proportional to\cite{lud:inf}
\[ (\ve_1-\ve_n)(\ve_2-\ve_{n-1})\ldots(\ve_q-\ve_{n-q+1})(0)
   \sum_{a' \ne b' \ne\ldots\ne q' =1}^{n-q}
(\ve_{a'}-\ve_n)(\ve_{b'}-\ve_{n-1})\ldots(\ve_{q'}-\ve_{n-q+1})(R). \]
We shall start by expanding this into $2^{2q}$ monomials:
\begin{itemize}
\item Choose $N$ operators from the LHS with index $\in [n-q+1, n]$, $0\le N\le q$. There are
$\sum_{N=0}^q \binom{q}{N}$ ways of doing this.

\item Choose $M$ matching operators on the RHS, $0\le M\le N$: $\sum_{M=0}^{N}\binom{N}{M}$ ways.

\item Choose $O$ other operators with index $\in [n-q+1, n]$ on the RHS, $0\le O\le q-N$: $\sum_{O=0}^{q-N}
\binom{q-N}{O}$ ways.
\end{itemize}
Note that $\sum_{N=0}^q \binom{q}{N}\sum_{M=0}^{N}\binom{N}{M}\sum_{O=0}^{q-N}\binom{q-N}{O}=2^{2q}$,
so we have accounted for all the terms.

\vspace{2ex}
For a typical monomial 
\[ \ve_1\ldots\ve_{q-N}\ve_q\ldots\ve_{q-N+1}\sum
\ve_{a'}\ldots\ve_{(q-M-O)'}\ve_q\ldots\ve_{q-M+1}\ve_{q-N}\ldots\ve_{q-N-O+1}, \]
the number of  explicit free indices $\ve_a' \ve_b'\ldots$ on
the RHS that are paired, $P$, runs over the range ${0 \le P \le {\rm Min}(q-N,q-M-O)}$. For a given $P$
we have the following factors:   
\begin{itemize}
\item A sign of $(-1)^{N+M+O}$
\item The number of ways of choosing the pairings, $\binom{q-N}{P}\binom{q-M-O}{P}P!$
\item $(q-M-O-P)$ explicit, unpaired free indices ranging over $(n-q)-(q-N)$ replicas, giving a factor of
 \[\frac{(n+N-2q)!}{(n+N-3q+M+O+P)!} \]
\item $(M+O)$ implicit free indices on the RHS (which must still be summed over) ranging over
$(n-q)-(q-M-O)$ replicas, a factor of
 \[ \frac{(n-2q+M+O)!}{(n-2q)!} \]
\end{itemize}

A term with $P$ pairings will produce a correlator 
\[C_{P+M,\,q-P-M}=\overline{\langle\ve(0)\rangle^{q-P-M}
\langle\ve(0)\ve(R)\rangle^{P+M}\langle\ve(R)\rangle^{q-P-M}} \]

Altogether, our decomposition into correlators becomes
\begin{multline*}
\langle\mathfrak{I}_{q,q}(0)\mathfrak{I}_{q,q}(R)\rangle=
\sum_{N=0}^q\binom{q}{N} \sum_{M=0}^{N}\binom{N}{M}\sum_{O=0}^{q-N}\binom{q-N}{O}
\sum_{P=0}^{{\rm Min}(q-N,\,q-M-O)}\!\!\!\!\!\!P!\,\binom{q-N}{P}\binom{q-M-O}{P} \\
(-1)^{N+M+O} \frac{(n+N-2q)!}{(n+N-3q+M+O+P)!}\frac{(n-2q+M+O)!}{(n-2q)!} C_{P+M,\,q-P-M}
\end{multline*}
For various $q$, this formula gives
\begin{eqnarray*}
q=1 & n(C_{1,0}-C_{0,1}) \\
q=2 & (n-1)(n-2)(C_{2,0}-2C_{1,1}+C_{0,2}) \\
q=3 & (n-2)(n-3)(n-4)(C_{3,0}-3C_{2,1}+3C_{1,2}-C_{0,3}) \\
q=4 & (n-3)(n-4)(n-5)(n-6)(C_{4,0}-4C_{3,1}+6C_{2,2}-4C_{1,3}+C_{0,4}) \\
q=5 & (n-4)(n-5)(n-6)(n-7)(n-8)(C_{5,0}-5C_{4,1}+10C_{3,2}-10C_{2,3}+5C_{1,4}-C_{0,5})
\end{eqnarray*}
i.e., the most antisymmetric irreps correspond to moments of the connected two-point correlation
function $\overline{G_c^q}$.

The above argument can easily be extended to irreps which are not the most antisymmetric. If we
consider now the irrep $\mathfrak{I}_{q,r}$ corresponding to Young tableau 

\hfil
\begin{picture}(185,35)
\put(0,15.5){\framebox(15,15){$\ve$}}
\put(15.5,15.5){\framebox(30,15){$\ldots$}}
\put(46,15.5){\framebox(15,15){$\ve$}}
\put(61.5,15.5){$\underbrace{
 \begin{picture}(61.5,15.5)
   \put(0,0){\framebox(15,15){$\ve$}}
   \put(15.5,0){\framebox(30,15){$\ldots$}}
   \put(46,0){\framebox(15,15){$\ve$}}
  \end{picture}
 }_{q-r}$}
\put(123,15.5){$\underbrace{\framebox(60,15){$\ldots$}}_{n-q-r}$}
\put(0,0){$\underbrace{
  \begin{picture}(60,15.5)
   \put(0,0){\framebox(15,15){}}
   \put(15.5,0){\framebox(30,15){$\ldots$}}
   \put(46,0){\framebox(15,15){}}
  \end{picture}
 }_r$}
\end{picture}
\hfil

\bigskip
the following changes occur:
\begin{itemize}
\item The $(q-M-O-P)$ explicitly unpaired indices now range over ${(n-r)-(q-N)}$
replicas, giving a factor of
\[\frac{(n+N-q-r)!}{(n+N-2q-r+M+O+P)!}\]
\item The $(M+O)$ implicit free indices range over ${(n-r)-(q-M-O)}$ replicas, a factor of
\[\frac{(n-q-r+M+O)!}{(n-q-r)!}\]
\end{itemize}

Incorporating these changes, we obtain (for instance)
\begin{eqnarray*}
q=1,r=0 & C_{1,0}+(n-1)C_{0,1} \\
q=2,r=0 & 2C_{2,0}+4(n-2)C_{1,1}+(n-2)(n-3)C_{0,2} \\
q=2,r=1 & n(C_{2,0}+(n-4)C_{1,1}+(3-n)C_{0,2})
\end{eqnarray*}

\end{document}